\documentclass[twocolumn]{aastex701}
\usepackage{amsmath}
\usepackage{multirow}
\newcommand{\oiiidoublet}{[O\,{\footnotesize III}]\,$\lambda\lambda4960,5008$}
\newcommand{\oiii}{[O\,{\footnotesize III}]\,$\lambda5008$}
\newcommand{\oiiileft}{[O\,{\footnotesize III}]\,$\lambda4960$}
\newcommand{\kms}{$\,\mathrm{km\,s^{-1}}$}

\begin{document}

\title{Source--Line Associations in JWST/NIRSpec Dense-Shutter Spectroscopy}

\author[0000-0002-8876-5248]{Zihao Wu}
\affiliation{Center for Astrophysics $|$ Harvard \& Smithsonian, 60 Garden St., Cambridge MA 02138 USA}
\email[show]{zihao.wu@cfa.harvard.edu}

\author[0000-0002-2929-3121]{Daniel J.\ Eisenstein}
\affiliation{Center for Astrophysics $|$ Harvard \& Smithsonian, 60 Garden St., Cambridge MA 02138 USA}
\email{deisenstein@cfa.harvard.edu}

\author[0000-0003-2388-8172]{Francesco D'Eugenio}
\affiliation{Kavli Institute for Cosmology, University of Cambridge, Madingley Road, Cambridge, CB3 0HA, UK}
\affiliation{Cavendish Laboratory, University of Cambridge, 19 JJ Thomson Avenue, Cambridge, CB3 0HE, UK}
\email{fd391@cam.ac.uk}

\author[0000-0002-7524-374X]{Erica Nelson}
\affiliation{Department for Astrophysical and Planetary Science, University of Colorado, Boulder, CO 80309, USA}
\email{Erica.June.Nelson@Colorado.edu}

\author[0000-0002-9280-7594]{Benjamin D.\ Johnson}
\affiliation{Center for Astrophysics $|$ Harvard \& Smithsonian, 60 Garden St., Cambridge MA 02138 USA}
\email{benjamin.johnson@cfa.harvard.edu}

\author[0000-0002-8651-9879]{Andrew J.\ Bunker}
\affiliation{Department of Physics, University of Oxford, Denys
 Wilkinson Building, Keble Road, Oxford OX1 3RH, UK}
\email{andy.bunker@physics.ox.ac.uk}

\author[0000-0001-7673-2257]{Zhiyuan Ji}
\affiliation{Steward Observatory, University of Arizona, 933 N. Cherry Avenue, Tucson, AZ 85721, USA}
\email{zhiyuanji@arizona.edu}

\author[0000-0002-5104-8245]{Pierluigi Rinaldi}
\affiliation{Department of Astronomy, The University of Texas at Austin, Austin, TX 78712, USA}
\affiliation{Cosmic Frontier Center, The University of Texas at Austin, Austin, TX 78712, USA}
\email{prinaldi@utexas.edu}

\author[0000-0002-4271-0364]{Brant Robertson}
\affiliation{Department of Astronomy and Astrophysics, University of California, Santa Cruz, 1156 High Street, Santa Cruz, CA 95064, USA}
\email{brant@ucsc.edu}

\author[0000-0002-8224-4505]{Sandro Tacchella}
\affiliation{Kavli Institute for Cosmology, University of Cambridge, Madingley Road, Cambridge, CB3 0HA, UK}
\affiliation{Cavendish Laboratory, University of Cambridge, 19 JJ Thomson Avenue, Cambridge, CB3 0HE, UK}
\email{st578@cam.ac.uk}

\author[0000-0001-9262-9997]{Christopher N.\ A.\ Willmer}
\affiliation{Steward Observatory, University of Arizona, 933 N. Cherry Avenue, Tucson, AZ 85721, USA}
\email{cnaw@as.arizona.edu}

\author[0000-0002-7595-121X]{Joris Witstok}
\affiliation{Cosmic Dawn Center (DAWN), Copenhagen, Denmark}
\affiliation{Niels Bohr Institute, University of Copenhagen, Jagtvej 128, DK-2200, Copenhagen, Denmark}
\email{joris.witstok@nbi.ku.dk}

\begin{abstract}
We demonstrate the feasibility of source--line association in JWST/NIRSpec dense-shutter spectroscopy, with $\sim$2000 sources observed simultaneously per pointing regardless of spectral overlap. While JWST wide-field slitless spectroscopy (WFSS) can disentangle overlapping spectra using two grisms with perpendicular dispersion directions, NIRSpec can achieve this using gratings with different dispersions. Compared with WFSS, NIRSpec dense-shutter spectroscopy offers substantially higher sensitivity and broader wavelength coverage. We develop a Bayesian framework that combines cross-dispersion offsets, wavelength consistency, and changes in relative line positions between gratings to jointly determine source associations and redshifts. We test the method using mock observations based on the observed source distribution and emission-line fluxes from the JWST Advanced Deep Extragalactic Survey (JADES), accounting for the NIRSpec optical model, MSA operability, measurement noise, false-positive detections, and bad pixels. Although the spectral traces overlap extensively, fewer than 0.5\% of the simulated emission lines are blended with another line on the detectors in NIRSpec grating spectroscopy. Applying our method to these simulations, we obtain redshifts, at an accuracy of 98\%, for the $\sim$1800 sources that have detectable emission lines, of which $\sim$1500 are classified as confident and are 99.9\% accurate. We further investigate how the association accuracy depends on the number of grating configurations. By combining high multiplexing, high sensitivity, and broad wavelength coverage, NIRSpec dense-shutter spectroscopy  enables efficient, highly complete spectroscopic surveys for the high-redshift Universe.
\end{abstract}


\section{Introduction}
\label{sec:intro}
Large spectroscopic surveys of galaxies have long been an observational challenge. The JWST/NIRSpec micro-shutter assembly (MSA; \citealt{Jakobsen2022, Ferruit2022}) enables simultaneous observations of a large number of galaxies \citep[e.g.,][]{Bunker2024JADES, DEugenio2025DR3}. However, to avoid spectral overlap, only a small fraction of the sources in the field of view can be observed at once.

In contrast, JWST NIRCam wide-field slitless spectroscopy \citep[WFSS;][]{Greene2017} disperses every source in the field. The challenge is that the spectral traces of different galaxies may overlap, making it ambiguous which source a line belongs to. NIRCam/WFSS observations can break this ambiguity using two grisms with perpendicular dispersion directions, changing the relative positions of the lines and thereby reducing the confusion \citep[e.g.,][]{Sun2023ApJSlitless, Sun2025SAPPHIRES}. However, compared with NIRSpec MSA observations, NIRCam/WFSS has narrower wavelength coverage and much shallower depth, because the full sky background falls on the detectors.

NIRSpec dense-shutter spectroscopy combines the advantages of both approaches \citep{DEugenio2026DarkHorse}. It applies to NIRSpec grating observations that target emission lines and, as in WFSS, allows the spectra to overlap. Unlike WFSS, however, it retains the wide wavelength coverage and high sensitivity of NIRSpec, because the shutters block most of the sky background.

\begin{figure*}
    \centering
    \includegraphics[width=\linewidth]{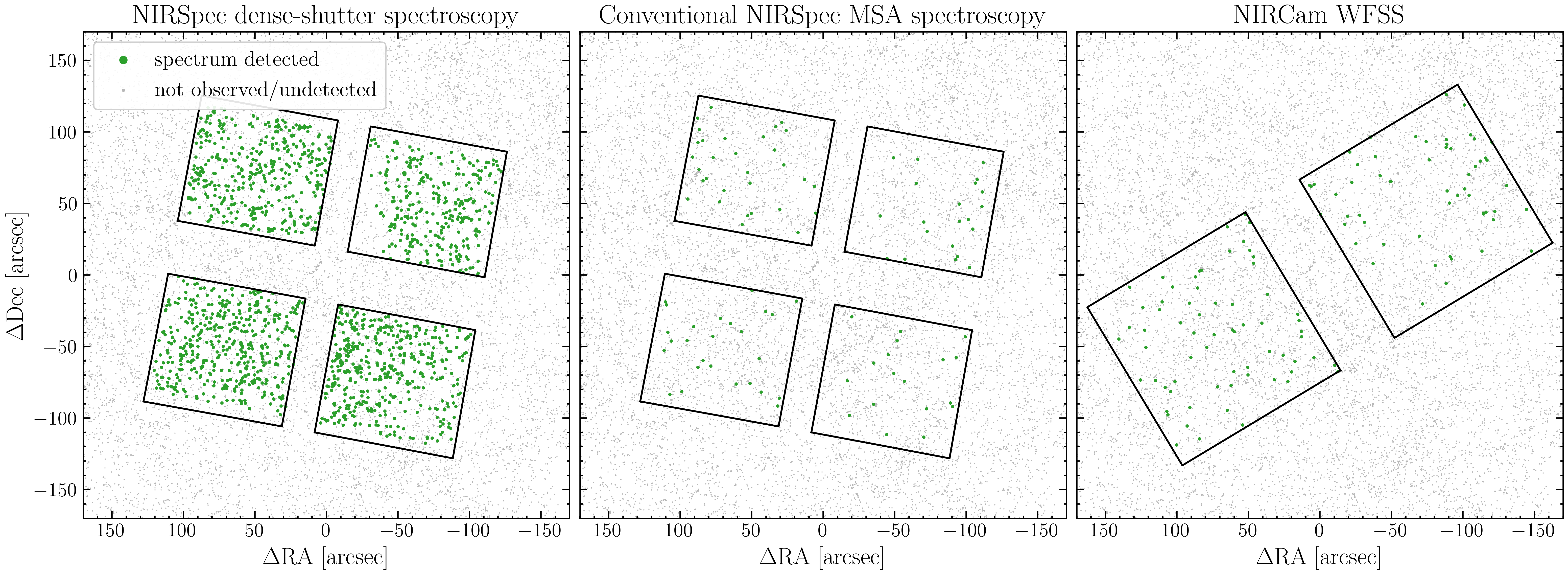}
    \caption{Sources with spectra obtained in three observing modes of the same field. Left: NIRSpec dense-shutter spectroscopy with G235M/F170LP, in which we open every operable shutter that contains a source. Middle: a conventional conflict-free MSA configuration. Right: NIRCam WFSS/F444W observation. Gray points are sources in the JADES NIRCam catalog with $z_{\rm phot}>1.5$, and green points are those whose spectrum is observed and shows at least one detected emission line, with their line fluxes simulated from the JADES NIRSpec catalog (Section~\ref{sec:sim_catalog}). For a 2 hr pointing, we adopt typical line sensitivities of $6\times10^{-19}\,\mathrm{erg\,s^{-1}\,cm^{-2}}$ for NIRSpec and $2\times10^{-18}\,\mathrm{erg\,s^{-1}\,cm^{-2}}$ for WFSS/F444W. Black outlines mark the four MSA quadrants in the left and middle panels, and the two NIRCam long-wavelength modules in the right panel.}
    \label{fig:comparison}
\end{figure*}

Dense-shutter spectroscopy increases the number of targets per pointing by an order of magnitude. Conventional grating observations usually target less than 200 sources, whereas \citet{DEugenio2026DarkHorse} observed $\sim$850 at once with dense-shutter spectroscopy and obtained robust redshifts for $\sim$540 galaxies. Even this falls short of maximum multiplexing: if every source in the open shutter area is observed, the MSA can place $\sim$2000 sources at $z = 0.1$--$10$ down to $m_{\rm F444W} = 29$\,mag in a single pointing (Figure~\ref{fig:comparison}; Section~\ref{sec:simulation}). It also largely removes the target preselection that the avoidance of spectral overlap would otherwise impose.

In the experience of \citet{DEugenio2026DarkHorse}, the main challenges of dense-shutter spectroscopy are source--line association and background subtraction. While background subtraction already performs well with traditional strategies, source--line association has so far been done only by visual inspection, and no general method for this problem exists. Nor has it been tested how much spectral overlap the association can tolerate while remaining reliable.

In most slitless grism situations, observers rely on rotation of the dispersion direction to associate lines with objects. Here, we highlight a different method, utilizing the fact that NIRSpec has gratings with different dispersions. Lines at different wavelengths move by different numbers of pixels, revealing accurate associations with the target objects. This method is particularly effective in dense-shutter spectroscopy because one can accomplish this motion without changing the position of the objects relative to the grid of the MSA, so that there is no change of the target list, as would happen if one rotated the telescope.

This work demonstrates the reliability of source--line association in dense-shutter spectroscopy, even in the densest case, where almost every operable shutter containing a source is opened. We develop a Bayesian framework that combines the changes in relative line positions between gratings with the wavelengths and the cross-dispersion positions of the lines, and that quantifies the confidence of each association. Using mock observations, we evaluate its accuracy and its dependence on the grating configurations.

This paper is organized as follows. Section~\ref{sec:simulation} describes our mock observations based on the JWST Advanced Deep Extragalactic Survey \citep[JADES;][]{Eisenstein2026JADES}. Section~\ref{sec:algorithm} presents a Bayesian method for source--line association. Section~\ref{sec:performance} shows the performance of the method on the mock observations. Section~\ref{sec:discussion} discusses the factors not considered in this work. Section~\ref{sec:advantages} describes the advantages of dense-shutter spectroscopy over other observing modes and its potential for future surveys, and Section~\ref{sec:summary} summarizes.

\section{Mock Observations}
\label{sec:simulation}

We describe our simulation of the emission-line distribution on the NIRSpec detectors. The mock observations account for the real source distribution of the JADES survey, the NIRSpec optics, and the actual MSA shutter positions and operability.

We simulate only the quantities required for source--line association: the position of each source in the MSA plane, and the two-dimensional position and count rate of each emission line on the detectors. We do not model the line profiles or the continuum emission. Our mock data should therefore be regarded as the products of line detection, namely the detector position and count rate of each line together with their uncertainties. We also include realistic artifacts, such as false-positive line detections and lines lost to bad pixels.

We assume an exposure of 2 hr in each of the four grating/filter configurations. According to the JWST Exposure Time Calculator, this reaches a 5$\sigma$ line-flux sensitivity of $\sim$$6\times10^{-19}\,\mathrm{erg\,s^{-1}\,cm^{-2}}$ for a centered point source with an unresolved line, in all four configurations and over most of their wavelength ranges.

\subsection{Source Catalog and Emission-Line Fluxes}
\label{sec:sim_catalog}

Our mock observations use the source distribution of the JADES NIRCam photometric and morphological catalogs \citep{Robertson2026, Carreira2026}. We select sources with cataloged F444W magnitudes between 25 and 29, measured in a circular aperture of $0\farcs15$ radius. Galaxies with $z_{\rm phot} < 1.5$ are randomly downsampled by a factor of four, because their H$\alpha$ falls outside the wavelength coverage of G235M/F170LP and they are therefore lower-priority targets. The true redshift of each selected source is drawn around its photometric redshift \citep{Hainline2026, Robertson2026}, from a Gaussian of scatter $0.03\,(1+z_{\rm phot})$ with a 5\% fraction of catastrophic outliers spread uniformly within $\pm2$ of $z_{\rm phot}$, truncated to $z > 0$.

We assign emission-line fluxes to these sources from the JADES NIRSpec catalog \citep{CurtisLake2026, scholtz_jades_2026}. Each NIRCam source inherits the line fluxes of the NIRSpec source with the closest F444W flux, selected among those whose redshift $z$ satisfies $|z_{\rm phot} - z| < 0.5$. We include fifteen strong lines: [\ion{O}{2}]$\lambda3727$, [\ion{Ne}{3}]$\lambda3868$, H$\gamma$, H$\beta$, [\ion{O}{3}]$\lambda4960$, [\ion{O}{3}]$\lambda5008$, [\ion{N}{2}]$\lambda6550$, H$\alpha$, [\ion{N}{2}]$\lambda6585$, [\ion{S}{2}]$\lambda6718$, [\ion{S}{2}]$\lambda6733$, [\ion{S}{3}]$\lambda9533$, \ion{He}{1}$\lambda10833$, Pa$\beta$, and Pa$\alpha$, adopting the cataloged flux of each line detected with SNR $> 4$. Lines missing from the catalog are filled in from their ratios to the detected lines, using the empirical median and scatter of each line ratio measured over the entire NIRSpec catalog (Figure~\ref{fig:line_ratios}). Finally, we scale the inherited fluxes by the ratio of the F444W flux of the NIRCam source to that of the NIRSpec source, so that the line fluxes track the continuum brightness of the NIRCam source, and add an independent 10\% random scatter to each line.

The galaxy continuum is below the detector noise, so we neglect its effect in the simulation. For a source of AB magnitude $m$, the continuum flux falling on one detector pixel along the dispersion direction is

\bigskip
\begin{equation}
\label{eq:continuum}
\begin{split}
F_{\rm cont, pix} &= f_\nu \frac{\, c}{\lambda\, R\, N_{\rm pix/res}} \\
&= 3.1\times10^{-21}\,\mathrm{erg\,s^{-1}\,cm^{-2}\,pixel^{-1}} \\
&\quad \times 10^{-0.4(m-29)}
\left(\frac{4\,\mu\mathrm{m}}{\lambda}\right)
\left(\frac{1000}{R}\right)
\end{split}
\end{equation}

\noindent
where $f_\nu$ is the flux density of the source continuum, $R$ is the resolving power, and $N_{\rm pix/res} = 2.2$ is the number of pixels per resolution element. Even for the brightest sources in our sample ($m = 25$), the continuum is only $1.2\times10^{-19}\,\mathrm{erg\,s^{-1}\,cm^{-2}\,pixel^{-1}}$, subdominant to the detector noise of a line measurement.

\begin{figure*}
\centering
\includegraphics[width=1\textwidth]{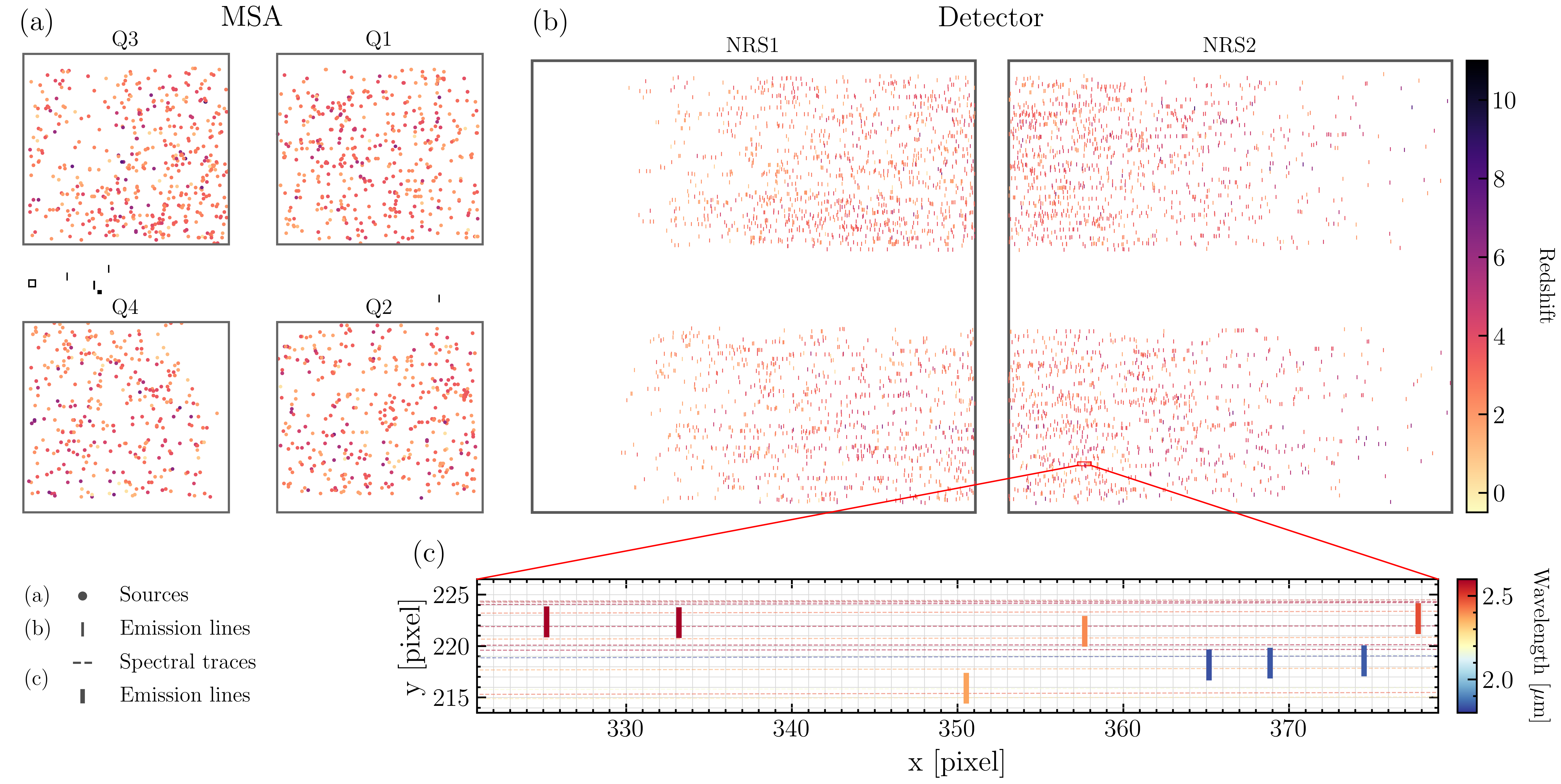}
\caption{Overview of a simulated maximally multiplexed NIRSpec observation in G235M/F170LP.
(a) Sources in open MSA shutters in the four quadrants, colored by redshift.
(b) Emission lines dispersed onto the two detectors, NRS1 and NRS2.
(c) Zoom-in on a small detector region: thick vertical bars are emission lines, colored by wavelength, and dashed curves are the spectral traces of all open shutters crossing this region.
Emission lines rarely blend on the detector even at maximum multiplexing.
\label{fig:overview}}
\end{figure*}

\bigskip
\subsection{Instrument Model}
\label{sec:sim_instrument}

We map each source from the sky onto the NIRSpec detectors in two steps: from the sky to the MSA plane, which determines the shutter of the source and how much of its flux enters that shutter, and from the MSA plane to the detectors, which determines where its emission lines land and how bright they are. We then add measurement errors and detector artifacts.

\subsubsection{From Sky to MSA}
\label{sec:sim_sky_to_msa}

As an example case, we simulate a pointing centered at (R.A., Dec.)\ $= (53.06,  -27.86)$ deg in GOODS-South, with the V3 position angle optimized to $31^\circ$ to maximize the number of sources in open shutters. We transform the R.A.\ and Dec.\ of each source into the JWST telescope frame (V2, V3), and then into MSA-plane coordinates through the NIRSpec geometric distortion model, following the algorithms of the eMPT \citep{Bonaventura2023eMPT}. Velocity aberration from the spacecraft motion is included as a first-order magnification of the field, accurate to better than 15 mas. Our implementation is publicly available\footnote{\url{https://github.com/zihaowu-astro/hMPT}} and has been validated against the MSA Planning Tool (MPT) in the Astronomer's Proposal Tool (APT).

We open a shutter for every source that meets the ``entire open shutter area'' criterion of the MPT, meaning that the centroid of the source falls anywhere inside the open area of an operable shutter. Each shutter has an open area of $0\farcs20 \times 0\farcs46$ on a pitch of $0\farcs27 \times 0\farcs53$, so the opaque bars cover 36\% of the MSA. Sources behind the bars, or in shutters that are inoperable in the current APT operability map, cannot be observed and are discarded. Typically $\sim$2000 sources remain in open shutters for a single pointing.

We model path loss considering the source size and the position of the source within the shutter. For simplicity, we model both the source and the point spread function (PSF) as circular Gaussians. The width of each source is set to its effective radius in the JADES data release~5 morphological catalog \citep{Carreira2026}, measured from the F200W image with \texttt{pysersic} \citep{Pasha2023pysersic}. We ignore the ellipticity for simplicity. The PSF at the MSA plane is fit to the \texttt{STPSF} model of the telescope and fore-optics, with $\sigma$ growing from $0\farcs014$ at $1\,\mu$m to $0\farcs068$ at $5\,\mu$m. The profile arriving at the MSA is the convolution of the two and is therefore wavelength dependent. For each source, we integrate this profile analytically over the open shutter area, centered on the intra-shutter position of the source, to obtain the in-shutter flux. Because the catalog fluxes are measured in a circular aperture of $0\farcs15$ radius, we normalize the in-shutter flux by the encircled energy of the same profile within that aperture. The resulting path-loss correction converts a cataloged aperture flux into the flux entering the shutter.

For every source in an open shutter, this step yields its MSA quadrant, its shutter indices within that quadrant, its offsets from the shutter center along the dispersion and cross-dispersion directions, and its path-loss correction as a function of wavelength. These quantities are the input to the MSA-to-detector model described next.

\begin{deluxetable}{lccc}
\tablecaption{Emission-line crowding on the NIRSpec detectors in the simulated maximum-multiplexing observation\label{tab:line_density}}
\tablehead{
  \colhead{Configuration} & \colhead{$N_{\rm line}$} &
  \colhead{Median Sep.} & \colhead{Blend Frac.} \\
  \colhead{} & \colhead{} & \colhead{(pixels)} & \colhead{}
}
\startdata
G235M/F170LP & 4696 & 14.5 & 0.43\% \\
G235H/F170LP & 3706 & 17.1 & 0.38\% \\
G395M/F290LP & 1281 & 26.6 & 0.00\% \\
G395H/F290LP & 1184 & 32.0 & 0.00\% \\
\enddata
\tablecomments{$N_{\rm line}$ is the number of emission lines above the detection limit in each grating/filter configuration (Section~\ref{sec:sim_noise}). The median separation is the median distance $\sqrt{\Delta x^{2} + \Delta y^{2}}$ from each line centroid to that of the nearest other line. The blend fraction is the fraction of lines whose nearest-line distance is smaller than 1 pixel.
Line blending due to two sources falling into the same shutter is not accounted for.}
\end{deluxetable}

\subsubsection{From MSA to Detector}
\label{sec:sim_msa_to_detector}

We compute the detector locations of the emission lines of each source with the NIRSpec optical model using the \texttt{msafit} software \citep{deGraaff2024msafit}. For each grating and filter, the model gives the spectral trace of a shutter across the two NIRSpec detectors, NRS1 and NRS2: the pixel position of the dispersed light as a function of wavelength. We evaluate the trace at the effective centroid of the source within the shutter.

We simulate four configurations: G235M/F170LP, G235H/F170LP, G395M/F290LP, and G395H/F290LP. For each, we adopt the full wavelength range transmitted by the optics: 1.7--5.5 $\mu$m for the F170LP filter and 2.9--5.5 $\mu$m for the F290LP filter. Our range for F170LP is wider than the nominal range of 1.7--3.1 $\mu$m, which is truncated to avoid contamination by second-order spectra \citep{Jakobsen2022}. We keep the full range because the optics remain transmissive beyond 3.1 $\mu$m, and because we simulate the second-order lines. For an individual source, some wavelengths may fall outside the detectors, so that only part of the wavelength range is available, depending on the location of its shutter.

We adopt the grating and filter transmission curves from the JWST User Documentation, and compute the intensity of each line as its flux multiplied by the shutter throughput of the source and by the grating and filter transmission at the observed wavelength. This intensity is proportional to the count rate of the line on the detector. For the second-order spectra, we adopt the response measured from NIRSpec IFU observations in G235M/F170LP \citep{Parlanti2026} and assume the same response for G235H/F170LP. The intensities of second-order lines are usually only $\sim$1\% of those of the first-order lines.

We add noise to the intensity and the position of each line to simulate measurement errors. The noise budget is dominated by the detector and is therefore independent of the line intensity \citep{DEugenio2026DarkHorse}, so we perturb the line intensities with Gaussian noise at the level of a typical 2 hr exposure. Although this noise is the same for every line, the corresponding sensitivity in flux is not: because the intensity already includes the shutter throughput of the source and the grating and filter transmissions at the observed wavelength, the limiting flux varies with wavelength and from source to source. The widths of a line along the dispersion and cross-dispersion directions are given by the standard deviation of the spatial profile of the source truncated by the shutter in each direction, broadened by one pixel to represent the spectrograph optics and the pixel sampling. These widths set the positional uncertainties adopted below. We do not consider velocity broadening.

For the position of a line, we adopt an error of its width divided by its SNR, with a floor of 0.2 pixel. In the cross-dispersion direction, we conservatively impose a further floor of one fifth of the line width, which accounts for a possible offset between the centroid of the line emission and the NIRCam photometric centroid from which we predict the trace. We then perturb the position of each line by Gaussian noise with these errors.

\subsubsection{Detector Artifacts}
\label{sec:sim_noise}

We inject false-positive lines at random detector positions to mimic unremoved cosmic rays and other artifacts. Their number in each configuration is drawn from a Poisson distribution with a mean number of 100, and their intensities from a log-uniform distribution in the range $6$--$1000\times10^{-19}\,\mathrm{erg\,s^{-1}\,cm^{-2}}$. These lines enter the mock catalog indistinguishably from the real ones, so our association does not know which lines are spurious.

We place 10,000 bad pixels at random positions on the two detectors, representing pixels masked because of cosmic-ray hits and other detector defects. A line falling within 2 pixels of a bad pixel is regarded as contaminated and removed from the catalog, so 1.5\% of the detector area is unusable. This is roughly the fraction that the cosmic-ray rate of \citet{Boker2023} leaves in a typical observation of our depth, once the standard data processing has recovered most of the affected pixels. Unlike the false positives, the positions of the bad pixels are recorded and available to the association, as they are in real observations.

\begin{figure*}
\centering
\includegraphics[width=\textwidth]{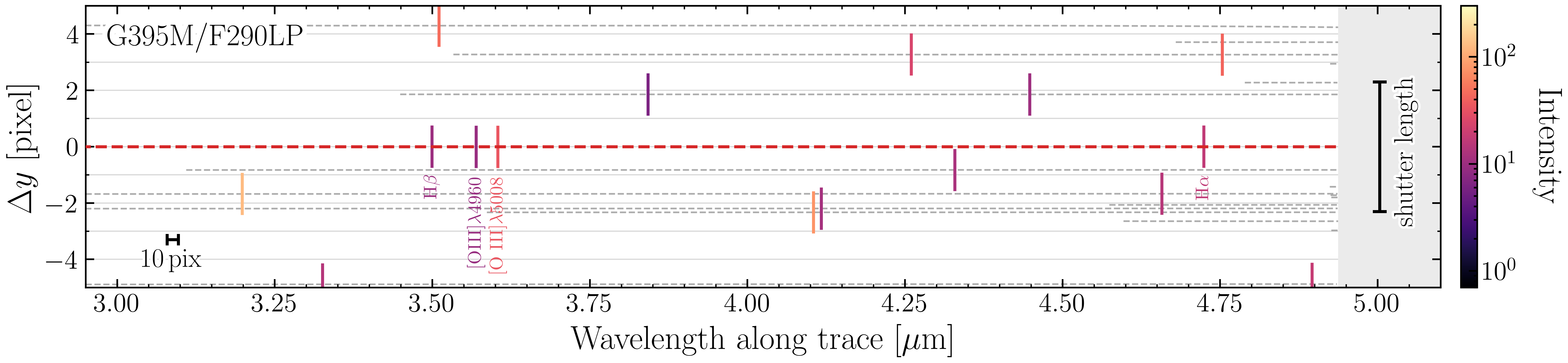}
\caption{Emission lines near the spectral trace of an example source in G395M/F290LP.
The red dashed line is the predicted trace of the target source ($\Delta y = 0$ by construction); gray dashed lines are the traces of other open shutters crossing this strip.
Vertical bars are detected emission lines, colored by intensity: the target's own lines (H$\beta$, \oiiidoublet, and H$\alpha$) fall on the predicted trace, while contaminating lines from other sources --- which would blend into a 1D extraction --- show measurable offsets in the cross-dispersion direction.
The shaded region marks wavelengths without detector coverage, and the vertical scale bar shows the shutter length.
\label{fig:trace}}
\end{figure*}

\begin{figure*}
\centering
\includegraphics[width=\textwidth]{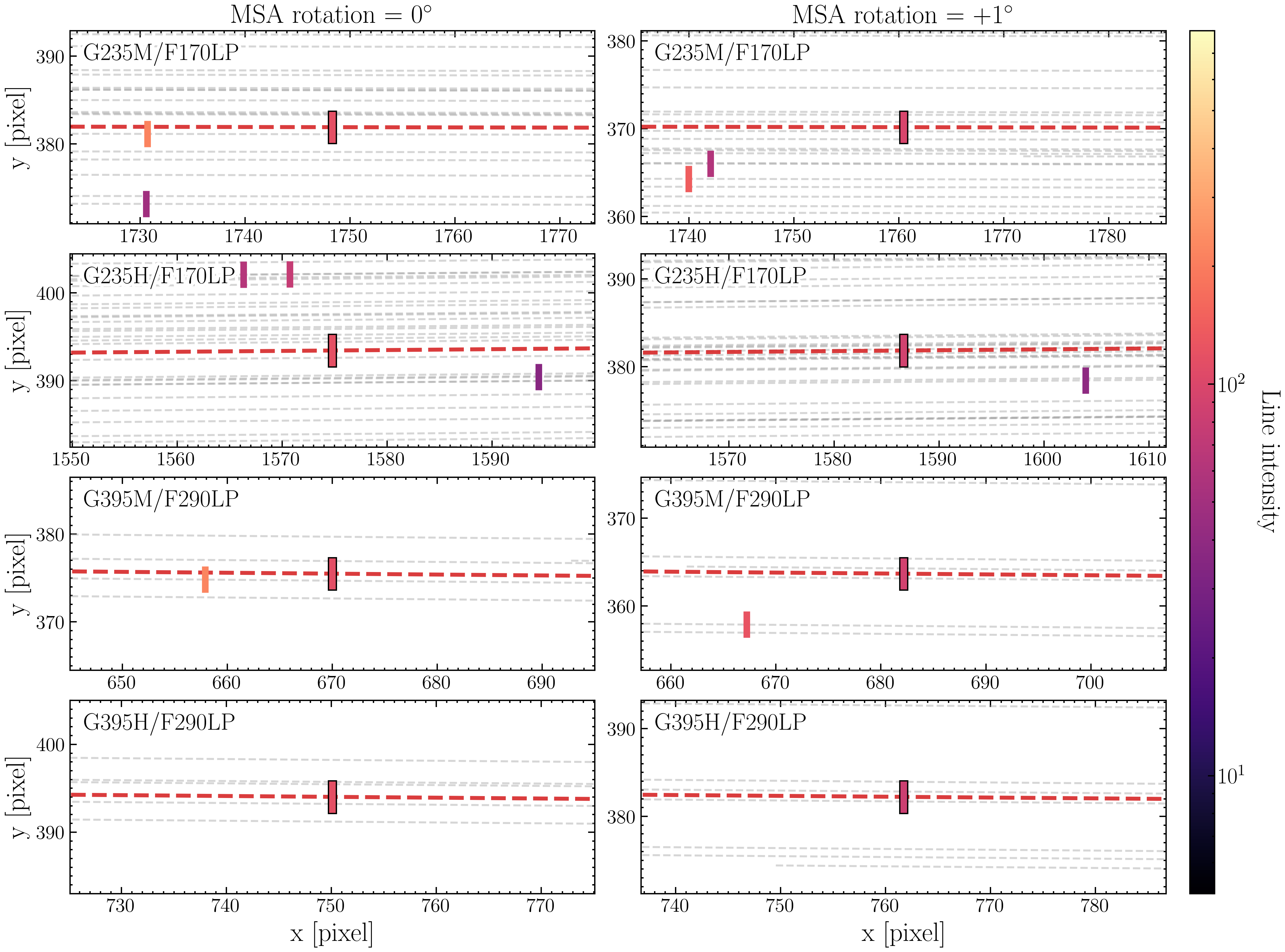}
\caption{The H$\alpha$ line of an example source observed in the four grating/filter configurations (rows) and two MSA pointings differing by a $1^\circ$ rotation (columns).
The red dashed line is the target's predicted trace, and the boxed line is the one associated with the target; gray dashed lines are traces of other sources.
The target line follows the predicted trace in every configuration, whereas nearby contaminating lines shift in both the dispersion and cross-dispersion directions across configurations, enabling robust source--line association.
\label{fig:cross_config}}
\end{figure*}

\subsection{Properties of Simulated Spectra}
\label{sec:sim_properties}

Figure~\ref{fig:overview} gives an overview of a simulated observation in G235M/F170LP, from the sources in open MSA shutters to their emission lines dispersed onto the detectors. Even in the crowded region of the zoom-in panel, the emission lines rarely blend: they are typically offset from one another along both the dispersion ($x$) and cross-dispersion ($y$) directions. 

Despite the high density of sources, the emission lines are sparsely distributed on the detectors and rarely blend (Table~\ref{tab:line_density}). Each observation contains 1,000--5,000 detected emission lines spread over the $8\times10^{6}$ pixels of the two detectors, and an unresolved line has a full width at half maximum of only $\sim$2 pixels. Even in G235M/F170LP, the most crowded configuration, the median separation between a line and its nearest neighbor is 14.5 pixels, about seven times the line width. Fewer than 0.5\% of the lines lie within 1 pixel of another line, so blending is usually negligible.

In contrast, the spectral traces from different sources may overlap.  However, their lines are usually offset in the cross-dispersion direction ($y$-direction), because the sources sit at different positions within their shutters (Figure~\ref{fig:trace}). At maximum multiplexing, we open a median of seven shutters per MSA row. Each shutter has a length corresponding to 4.6 pixels ($0\farcs46$) in the $y$-direction, and line centroids are usually measured to better than 1 pixel, so more than four sources per row can be separated by their offsets alone. The overlap is even lower because the spectral trace of a medium-resolution grating is shorter than the combined length of the two detectors. Additionally, line ambiguities can be further resolved using the wavelengths.

The relative positions of the lines also change from one configuration to another, and from one position angle to another (Figure~\ref{fig:cross_config}). Two sources in different shutters are a fixed distance apart along the dispersion direction, but the wavelength interval that this distance corresponds to depends on the dispersion of the grating. A line from a neighboring source therefore implies a different wavelength in each grating: it may match an emission line of the source in one grating but not in another. Only the true lines of a source have wavelengths that match in every configuration. Such information is powerful in determining source--line associations.

\begin{figure}
\centering
\includegraphics[width=0.48\textwidth]{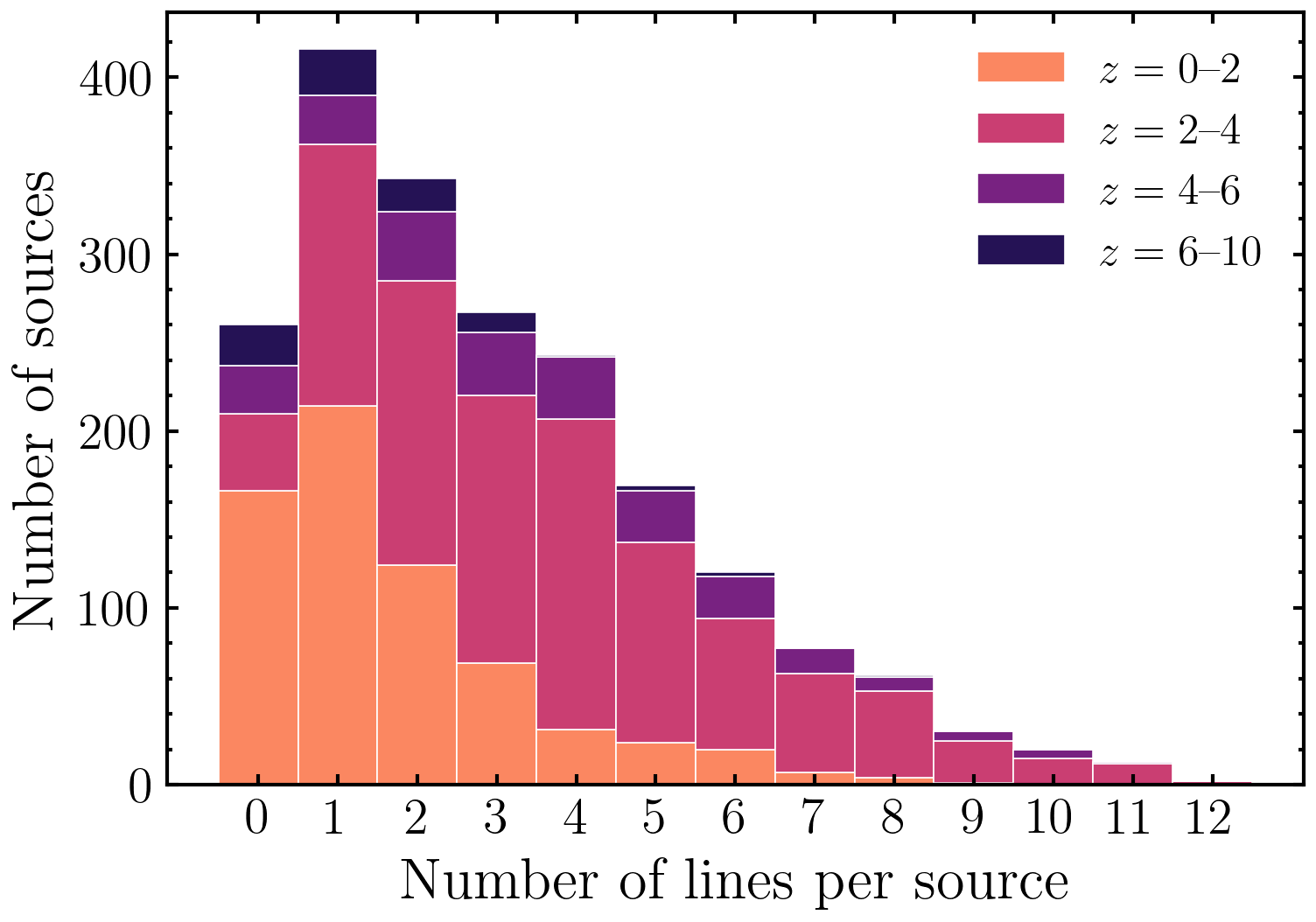}
\caption{Number of unique emission lines detected per source in the simulated maximum-multiplexing observation, with the four grating/filter configurations combined. A line counts once per source however many configurations or diffraction orders it is detected in. Only lines above the detection limit are counted, and the injected false-positive lines are excluded (Section~\ref{sec:sim_noise}). 
\label{fig:number}}
\end{figure}

\section{Source--Line Association}
\label{sec:algorithm}

We now describe a method for source--line association based on a Bayesian framework that evaluates the association probabilities.  For every source, our algorithm gives the five most probable sets of lines, the redshift each set implies, and its probability.

Our association uses four kinds of information, each of which becomes a term of the posterior odds in Sections~\ref{sec:algo_scoring}--\ref{sec:algo_global}.

\begin{enumerate}
    \item \emph{Wavelength concordance.} Two or more lines of one source must be explained by a single redshift within their wavelength uncertainties. Each additional concordant line adds to the evidence.

    \item \emph{Cross-dispersion position.} The line position must be consistent with the spectral trace of its source in the cross-dispersion direction (Figure~\ref{fig:trace}).

    \item \emph{Consistency between gratings.} A true line should reappear in every configuration with the depth and the wavelength coverage to detect it, whereas a line from a neighboring shutter generally shifts away from the predicted position when the dispersion changes (Figure~\ref{fig:cross_config}). 

    \item \emph{Physical prior.} We consider one physical prior on the line fluxes: we penalize solutions if a strong line is expected but missing, given the other detected lines (Section~\ref{sec:algo_priors}). This tells us, for example, whether a single detected line is \oiii\ or the three-times-fainter \oiiileft, which should not appear on its own. 
\end{enumerate}

None of these expectations is treated as a strict rule. We evaluate them probabilistically, each with an outlier fraction that keeps a single inconsistency from ruling out a solution by itself. For every source we enumerate all possible associations of lines and the redshifts that each of them implies, and we score them by their Bayesian evidence (Section~\ref{sec:algo_scoring}) and by penalties (Section~\ref{sec:algo_priors}). These candidate solutions form a conflict graph, because a line can belong to at most one source and a source can have only one redshift. We solve this graph in Section~\ref{sec:algo_global} and obtain the set of mutually compatible solutions with the highest total likelihood.

For sources with two or more lines, we use only the spectroscopic information listed above. For a source with a single line, however, the wavelength of that line cannot determine the redshift uniquely. We therefore apply a photometric redshift prior to the sources whose solution contains one line or none (Section~\ref{sec:algo_fewlines}).

For every source we further solve for its five most probable solutions and evaluate each of  their probabilities (Section~\ref{sec:algo_confidence}). We regard an association as confident when one solution dominates with $P(z) \geq 0.99$.

\subsection{Posterior Odds of a Candidate Solution}
\label{sec:algo_scoring}

We enumerate every possible association of lines with a source and evaluate its likelihood. We refer to each of them as a candidate solution: a set of detected lines explained as one source at a single redshift. For each source we consider all lines within $5\sigma$ in $\Delta y$ of its predicted trace (Figure~\ref{fig:trace}). We then consider every subset of these lines, and every redshift at which the lines of the subset agree in wavelength within $5\sigma$. We do not resolve conflicts between sources at this stage.

We score a solution $S$ by its log posterior odds against the null hypothesis that all of its lines are contaminants,

\begin{equation}
\label{eq:logodds}
\ln \mathcal{O}(S) = \sum_{i \in S} \ln \mathrm{BF}_i
+ \ln \Omega
+ \sum_{m} \ln P_{\mathrm{miss},m} ,
\end{equation}

\noindent
where $\mathrm{BF}_i$ is the Bayes factor of line $i$ (Equation~\ref{eq:bf}), $\Omega$ is the Occam factor that accounts for the search over redshift (Equation~\ref{eq:occam}), and $P_{\mathrm{miss},m}$ is the probability that an expected line escapes detection (Section~\ref{sec:algo_priors}). Log odds of zero mean that the solution is as probable as the null hypothesis, and a solution with negative log odds is less probable than having no line at all (Section~\ref{sec:algo_global}).

The Bayes factor weighs the probability that line $i$ belongs to the source against the probability that it is a contaminant. Assuming Gaussian measurement errors on the line position, it is

\begin{equation}
\label{eq:bf}
\ln \mathrm{BF}_i = \ln \mathcal{N}(\Delta\lambda_i;\, \sigma_{\lambda,i}) + \ln \mathcal{N}(\Delta y_i;\, \sigma_{y,i}) - \ln \rho_c ,
\end{equation}

\noindent
where $\mathcal{N}(x;\sigma)$ is a Gaussian of width $\sigma$ centered on zero, $\Delta\lambda_i$ is the offset of the line from the wavelength that the solution implies, and $\Delta y_i$ is its offset from the predicted trace, with the measurement uncertainties $\sigma_{\lambda,i}$ and $\sigma_{y,i}$ of Section~\ref{sec:sim_msa_to_detector}. We measure $\Delta\lambda$ in $\mu$m and $\Delta y$ in pixels, and the contaminant density $\rho_c$ per $\mu$m per pixel, so that the Bayes factor is dimensionless. We estimate $\rho_c$ from the number of false-positive lines per exposure (Section~\ref{sec:sim_noise}). We do not include the lines of the other sources in $\rho_c$, because most of them are eventually assigned to their own sources.

We first build the solutions separately in each grating observation, and combine them when their redshifts $\hat z$ agree within the uncertainties, where $\hat z$ and its uncertainty follow from the inverse-variance weighted mean of the redshifts implied by the individual lines. As the gratings are independent data sets, their log odds add after combining.

We include an Occam factor for the redshift search. Because the redshift is a free parameter, the wavelength match of a single-line source should not add any evidence. For a Gaussian wavelength likelihood, the Occam factor is given by the Laplace approximation,

\begin{equation}
\label{eq:occam}
\ln \Omega = \tfrac{1}{2}\ln\!\left(2\pi\sigma_z^2\right) + \ln p(\hat z),
\end{equation}

\noindent
where $\sigma_z$ is the redshift uncertainty, and $p(z)$ is the redshift prior. We adopt $p(z)$ as the global distribution of photometric redshifts with the same magnitude selection as our mock catalog (Section~\ref{sec:sim_catalog}). For a single line, $\sigma_z = \sigma_\lambda/\lambda_{\rm rest}$, so the first term cancels the wavelength term of Equation~\ref{eq:bf} up to a constant.

We treat second-order lines as follows. Where the grating has second-order transmission, we consider both the nominal first-order interpretation at $\lambda_{\rm app}$ and a second-order interpretation at $\lambda_{\rm app}/2$, correcting the flux of the latter for the second-order transmission. We then search for the corresponding first-order parent line of the same source. If the parent line is not detected although its wavelength is covered, we reject the second-order interpretation. If instead the parent line appears at the expected wavelength and with the expected flux, we accept the second-order identification and lock both lines to the source. Ambiguous cases, in which the parent wavelength falls outside the coverage, are retained and resolved by the global redshift inference.

\subsection{Inconsistency Penalty}
\label{sec:algo_priors}

We consider two  penalties for the non-detection of expected lines. The first applies to a line detected in one grating that should also appear in another but does not. The second applies to a source in which a line expected to be strong from the line ratios is not detected.

A solution that claims \oiiileft, for instance, should also show the brighter \oiii. A source with \oiii\ and [\ion{S}{3}]$\lambda9533$ should also show H$\alpha$, which lies between the two in wavelength and is usually brighter than [\ion{S}{3}]. Visual inspection relies on arguments of this kind to accept or reject a redshift, and we make them quantitative by evaluating the probability of a non-detection. Lines expected to fall in a detector gap or on a bad pixel are exempt.

The probability that a line escapes detection is

\begin{equation}
\label{eq:miss}
\ln P_{\mathrm{miss}} = \ln\!\left[\varepsilon + (1-\varepsilon)\,\Phi\!\left(t - \mathrm{SNR}_{\mathrm{exp}}\right)\right],
\end{equation}

\noindent
where $\Phi$ is the standard normal cumulative distribution, $t=5$ is the SNR threshold of detection, and $\mathrm{SNR}_{\mathrm{exp}}$ is the SNR expected where the line is missing. The outlier fraction $\varepsilon$ sets a floor on the penalty. We adopt $\varepsilon = 10^{-3}$ for a line missing in another grating, which accounts for unrecorded bad pixels in real observations, and $\varepsilon = 0.1$ for a line expected from the line ratios, which accounts for sources with unusual ratios, matching the typical rate measured in the JADES DR4 catalog. In the first case we compute $\mathrm{SNR}_{\mathrm{exp}}$ from the flux  in the gratings where the line is detected, corrected for the transmission.  In the second case we predict the flux from the empirical line ratios (Appendix~\ref{app:line_ratios}), accounting for  wavelength-dependent transmissions. We draw the line ratios in a Monte Carlo simulation and average the resulting probability of non-detection. We apply all these penalties only when $\mathrm{SNR}_{\mathrm{exp}} > 7$, because a line expected near the $5\sigma$ detection limit can easily fall below it owing to noise. We use a wide $10\sigma_\lambda$ window for the first penalty, to account for possible errors of the wavelength calibration between gratings in real observations.

\subsection{Resolving Conflicts}
\label{sec:algo_global}

The solutions of different sources may compete for the same lines. We resolve these conflicts by maximizing the total likelihood of all sources, subject to two conditions: each source has one redshift, and each line belongs to at most one source.

To speed up the optimization, we first lock the lines whose association is unambiguous on geometric grounds alone. We lock lines in two cases: (a) a source with more than four lines that are consistent in wavelength, $y$ position, and flux ratio, in which case all of these lines are locked to that source; and (b) a line that appears in at least two gratings with consistent wavelength, $y$ position, and flux, in which case these repeated detections are locked to that source. We require a strict consistency criterion of $<$1$\sigma$. In total, 4,041 of the 10,867 lines in the four gratings are locked. Seven of them  are locked to the wrong source by chance alignment, three from case (a) and four from case (b). We consider this error minor and accept it for the gain in computational efficiency. This locking is a computational shortcut applied before the global solve; it is a property of individual lines, not a statement about the confidence of a source's redshift (Section~\ref{sec:algo_confidence}).

The remaining solutions form a conflict graph, in which each candidate solution is a node weighted by its log odds (Equation~\ref{eq:logodds}), and two nodes are linked whenever they cannot both hold, either because they belong to the same source or because they claim the same line. What we seek is then a maximum-weight independent set: the set of mutually compatible solutions with the highest total log odds. We first decompose the graph into independent subgraphs, since the spectral traces of sources on different MSA rows rarely overlap and  conflict. In each subgraph we assign one boolean variable per candidate solution and maximize the total log odds. The problem is NP-hard in general, but our subgraphs are small enough to be solved exactly with the CP-SAT solver of Google OR-Tools\footnote{\url{https://developers.google.com/optimization}}, and the subgraph solutions together give the assignment of the whole field with the highest posterior probability. A line claimed by several sources is thereby awarded to the source whose total evidence suffers most from losing it.

\begin{figure*}
\centering
\includegraphics[width=0.9\textwidth]{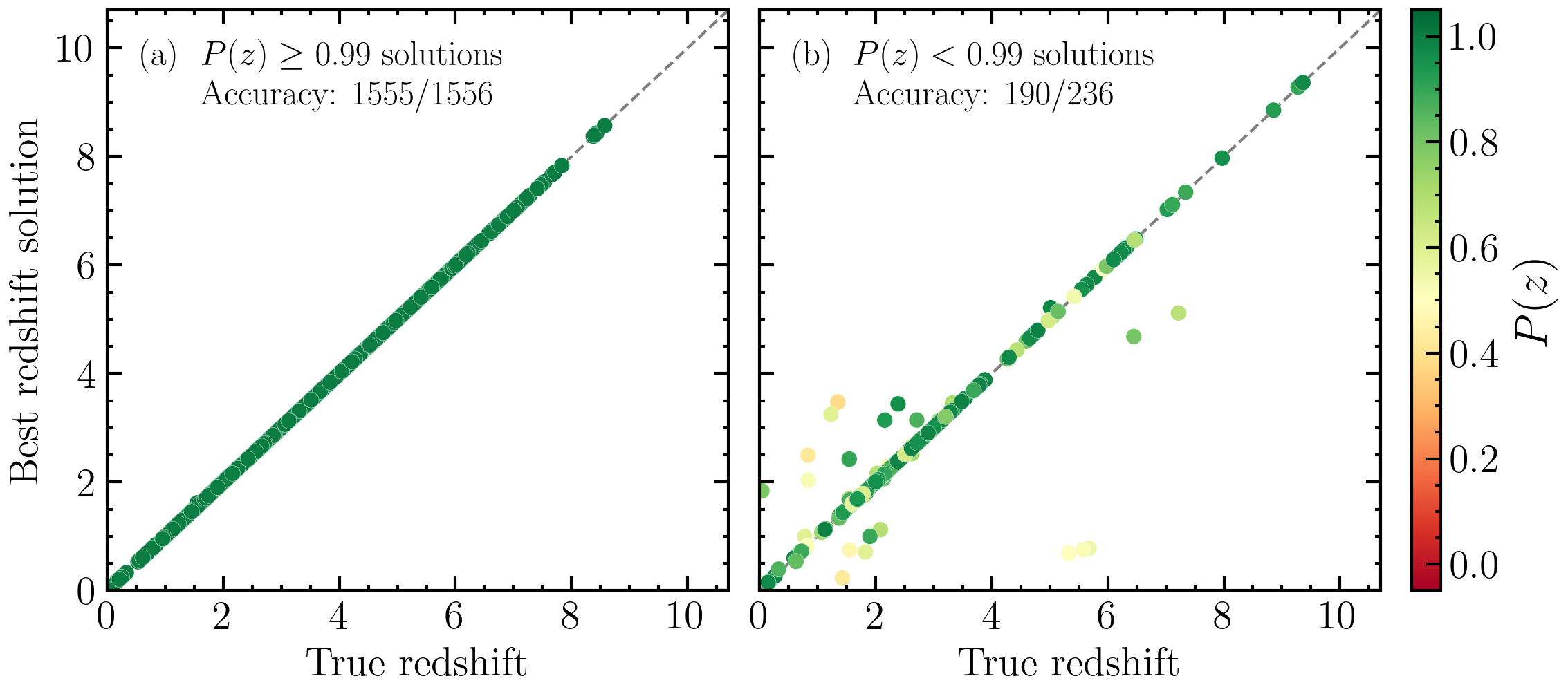}
\caption{Estimated versus true redshift for the simulated catalog, combining all four grating/filter configurations and including realistic noise and 100 injected false-positive lines. Left: sources with confident redshifts ($P(z) \geq 0.99$); right: the remaining sources. Points are colored by $P(z)$, the posterior probability of the best solution normalized over each source's top five solutions. Open circles mark sources whose multiple solutions share the same line association, typically because only a single line is detected: the redshift is then degenerate among line identifications, and we count the association as correct as long as the line is assigned to the correct source. In the left panel, the one incorrect confident solution lies near the correct redshift at $z \approx 1.5$; its line association is correct, but its [\ion{S}{2}]$\lambda6733$ line is misidentified as H$\alpha$ (Section~\ref{sec:perf_all}).
\label{fig:performance}}
\end{figure*}

\subsection{Sources with at Most One Line}
\label{sec:algo_fewlines}

A source with a single detectable line cannot be placed at a redshift by that line alone. However well its wavelength is measured, the line can be interpreted as any of the reference lines, each at a different redshift, and these identifications are degenerate. In our simulation, about 700 of the $\sim$2000 sources have one detectable line or none (Figure~\ref{fig:number}). For them the degeneracy can only be broken with the photometric redshift.

We apply the photometric prior only to the sources left with one or no line after a first round of the association. That round uses the spectroscopic evidence alone in the global optimization of Section~\ref{sec:algo_global}. For these sources we replace the redshift prior $p(z)$ of Equation~\ref{eq:occam} by $p(z \mid z_{\rm phot})$, which has a Gaussian core of width $0.03\,(1+z)$ and a 5\% fraction of uniform outliers, the same distribution from which the mock redshifts are drawn (Section~\ref{sec:sim_catalog}). We then reevaluate each redshift solution and solve the association again.

\subsection{Posterior Redshift Probabilities}
\label{sec:algo_confidence}

For each source we solve for the five most probable solutions and obtain a probability for each of them. The global assignment of Section~\ref{sec:algo_global} gives only the single best solution per source, so we re-solve its subgraph with the source forced to a different solution each time, excluding the solutions already found. We also include the null hypothesis that the source has no identification, with log odds of zero. We denote by $\Delta_k$ the total log odds of the subgraph when the source takes solution $k$, minus that of the global optimum, so $\Delta_k$ measures the evidence that the field as a whole loses under this solution. We convert these relative log odds into normalized probabilities using the softmax function,

\begin{equation}
\label{eq:pz}
P_k(z) =
\frac{e^{\Delta_k}}
{\sum_{j} e^{\Delta_j}},
\end{equation}

\noindent
where the sum in the denominator runs over the top five solutions. For each source, we regard its redshift as confident when $P(z) \geq 0.99$. We stress that $P(z)$ is not a per-source goodness of fit, but a softmax over the evidence evaluated jointly with the sources that compete for the same lines. A solution has $P(z) \geq 0.99$ only when  no alternative interpretation of its own lines, and no competing assignment of the lines among its neighbors, can reach comparable evidence.

\begin{figure*}
\centering
\includegraphics[width=0.93\textwidth]{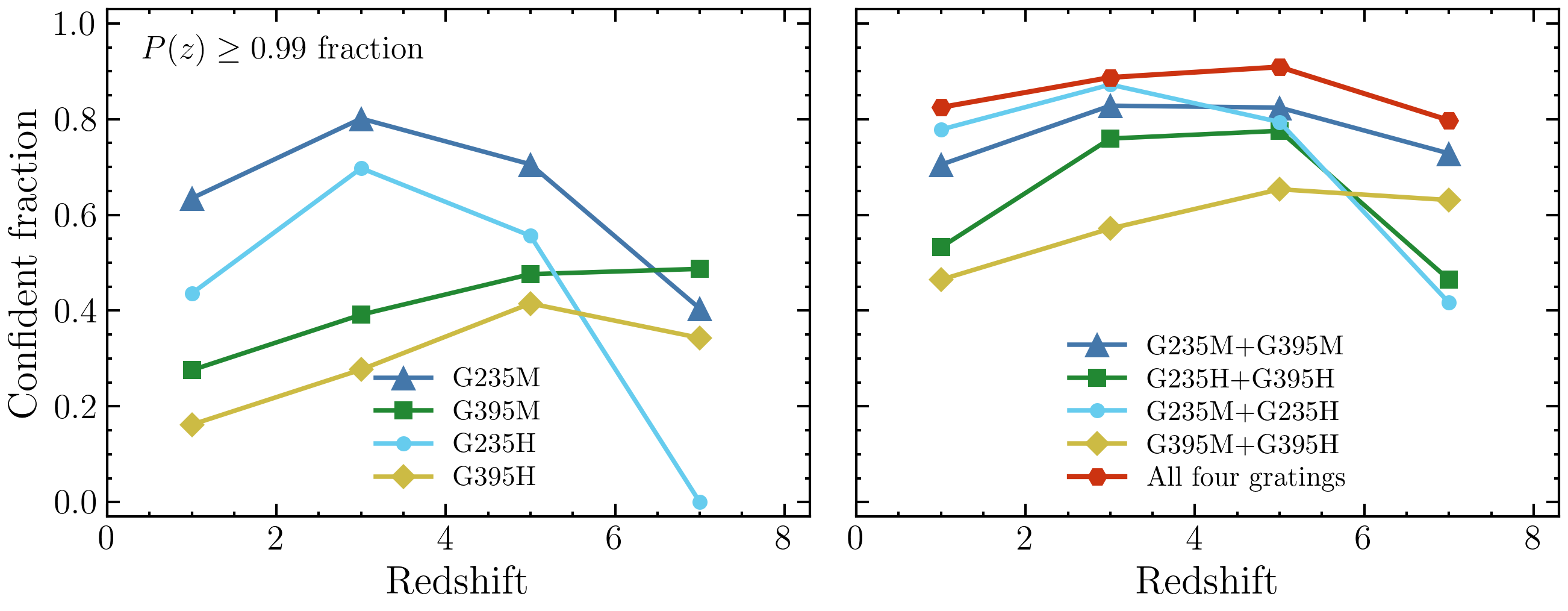}
\caption{Fraction of sources with confident redshift identifications ($P(z) \geq 0.99$; Section~\ref{sec:algo_confidence}) as a function of true redshift, for simulated observations at maximum multiplexing, $\sim$2000 sources per pointing. The fraction is computed among the sources that have at least one emission line above the detection limit in the gratings of the configuration shown. The numerator counts the sources whose adopted solution is both confident and correct.
Left: observations with a single grating/filter configuration.
Right: combinations of two configurations and of all four.
Sources are binned by true redshift in intervals of $\Delta z = 2$, plotted at $z = 1$, 3, and 5, while all sources at $z = 6$--$10$ are pooled into the last bin and plotted at $z = 7$.
\label{fig:confident_fraction}}
\end{figure*}

\begin{figure*}
\centering
\includegraphics[width=0.93\textwidth]{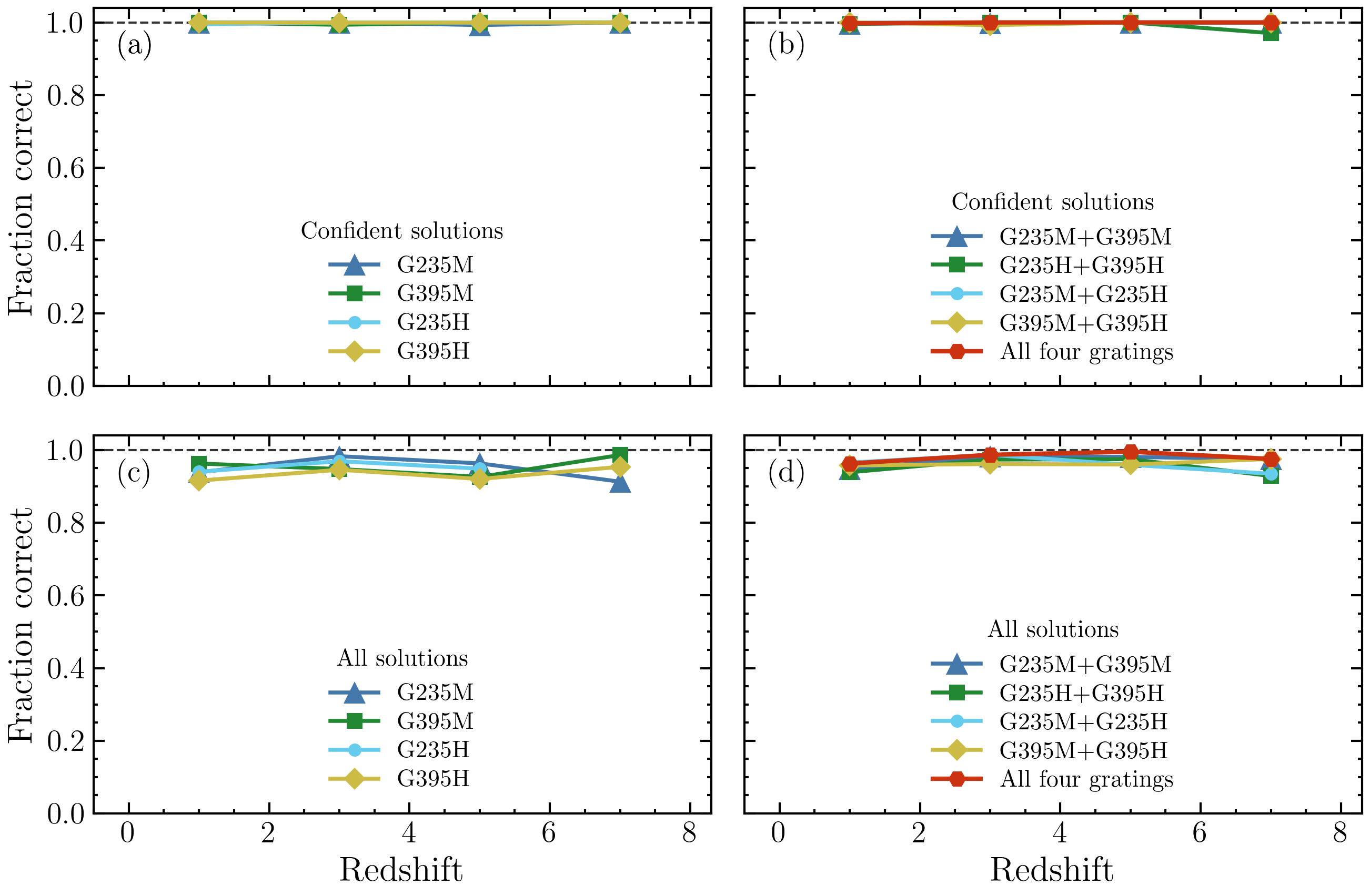}
\caption{Fraction of correctly identified redshifts as a function of true redshift, for simulated observations at maximum multiplexing, $\sim$2000 sources per pointing.
Panels (a) and (b) include only confident solutions ($P(z) \geq 0.99$; Section~\ref{sec:algo_confidence}); panels (c) and (d) include all solutions.
The left panels show observations with a single grating/filter configuration; the right panels show combinations of two configurations and of all four.
Sources are binned by true redshift in intervals of $\Delta z = 2$, plotted at $z = 1$, 3, and 5, while all sources at $z = 6$--$10$ are pooled into the last bin and plotted at $z = 7$.
\label{fig:accuracy}}
\end{figure*}

\section{Performance}
\label{sec:performance}

We apply the association to the mock observation of Section~\ref{sec:simulation}, which has 2067 sources in open shutters. Of these sources, 266 have no line above the detection limit in any grating, 435 have one line, and 1366 have two or more lines (Figure~\ref{fig:number}). We call an identification correct when its redshift agrees with the true value to $|\Delta z| \le 0.002$, and confident when $P(z) \ge 0.99$ (Section~\ref{sec:algo_confidence}).

\subsection{Redshift Recovery with Four Configurations}
\label{sec:perf_all}

Figure~\ref{fig:performance} compares the recovered and true redshifts. The association identifies redshifts for 1792 of the 2067 sources and none for the remaining 275. Of the 1556 confident identifications, 1555 are correct, as are 190 of the 236 with $P(z) < 0.99$. The accuracy is even higher at $z > 3$: 100\% for $P(z) \ge 0.99$, 99.9\% for $P(z) \ge 0.9$, and 98.9\% for all identifications.

One of the confident solutions is wrong because a single [\ion{S}{2}]$\lambda6733$ line is read as H$\alpha$, the true H$\alpha$ of that source falling just blueward of the F170LP cut-on. The line happens to be unusually strong in our simulation, and the two interpretations are penalized very differently. Read as [\ion{S}{2}]$\lambda6733$, it implies that [\ion{S}{3}], \ion{He}{1}, and the Paschen lines should all be detected redward of it. None is, and the resulting non-detection penalty (Section~\ref{sec:algo_priors}) suppresses the correct solution. Read as H$\alpha$, one of the brightest lines, it implies nothing else that should have been detected, so that interpretation escapes the penalty and carries the higher probability. 

We also recover most of the sources that have no emission line. We identify 275 sources as having no detectable line, of which 254 really have no detectable line. The other 21 have a single detectable line, which noise has displaced in the cross-dispersion direction, so that the zero-line hypothesis beats the one-line solution. Conversely, 12 of the sources with no detectable line receive a redshift from a contaminating line whose identification happens to fall in the core of the photometric prior, although none of them is confident. In all of these cases the correct solution is still among the five candidates, only with a slightly lower probability.

The accuracy primarily depends on the number of detectable lines. All 1032 sources with three or more lines are identified correctly, 1030 of them confidently. Of the 334 sources with two lines, 330 are correct and 310 confident. The four errors, all with $P(z) < 0.99$, are discussed in Section~\ref{sec:discussion_failures}. Of the 435 single-line sources, 414 are identified, 388 correctly, and 218 confidently, one of which is wrong.

\begin{deluxetable}{crrrrr}
\tablecaption{Redshift identification performance for different NIRSpec grating configurations.\label{tab:performance}}
\setlength{\tabcolsep}{10pt}
\tablehead{
\colhead{Configuration} &
\colhead{$N_{\rm det}$} &
\colhead{$N_{\rm solve}$} &
\colhead{$N_{\rm conf.}$} &
\colhead{Correct (conf.)} &
\colhead{Correct (all)}
}
\startdata
G235M       & 1638 & 1609 & 1192 & 0.999 & 0.967 \\
G395M       &  824 &  748 &  320 & 0.997 & 0.949 \\
G235H       & 1561 & 1491 &  936 & 0.999 & 0.958 \\
G395H       &  777 &  680 &  224 & 1.000 & 0.935 \\
G235M+G395M & 1711 & 1682 & 1344 & 0.999 & 0.973 \\
G235H+G395H & 1702 & 1651 & 1161 & 0.998 & 0.963 \\
G235M+G235H & 1726 & 1699 & 1415 & 1.000 & 0.977 \\
G395M+G395H &  931 &  892 &  529 & 0.996 & 0.962 \\
All four gratings & 1801 & 1792 & 1556 & 0.999 & 0.981
\enddata
\tablecomments{$N_{\rm det}$ is the number of sources with at least one emission line above the detection limit in the gratings used, out of 2067 sources in open shutters. $N_{\rm solve}$ is the number of these sources for which the pipeline associates at least one emission line, i.e., whose best solution is not the null hypothesis, and $N_{\rm conf.}$ the number with a confident redshift ($P(z) \ge 0.99$; Section~\ref{sec:algo_confidence}). The last two columns give the fraction of correct redshifts ($|\Delta z| \le 0.002$) among the confident and among all solved sources.}
\end{deluxetable}

\subsection{Dependence on Grating Configurations}
\label{sec:perf_configs}

We repeat the association with subsets of the configurations: each grating alone, two gratings combined, and all four combined (Figures~\ref{fig:confident_fraction} and \ref{fig:accuracy}). The exposure time in each grating is fixed at 2 hr (Section~\ref{sec:simulation}), so the comparison is at fixed depth per grating rather than at fixed total exposure time. A source without a detectable line in the gratings used cannot be identified there, so we quote the confident fraction both relative to the sources with at least one detectable line in those gratings and relative to all sources.

Figure~\ref{fig:confident_fraction} shows how the confident fraction depends on redshift, following the wavelength coverage of each grating/filter configuration. G235M/F170LP and G235H/F170LP peak at $z = 2$--$4$, where H$\beta$, \oiiidoublet, and H$\alpha$ are all covered. At $z > 6$, \oiii\ moves beyond the detector coverage of G235H, which then identifies no source confidently, and H$\alpha$ leaves the G235M range at $z \simeq 7.4$. G395M/F290LP and G395H/F290LP instead improve with redshift, as H$\alpha$ and then \oiii\ enter their range. Below $z \simeq 1.6$, H$\alpha$ is bluer than the F170LP cut-on, and the identifications can only rely on faint [\ion{S}{3}], \ion{He}{1}, and the Paschen lines, making the redshift association less reliable (Figure~\ref{fig:number}).

Combining configurations improves the confident fraction substantially over any single grating, for three reasons. First, the changes in dispersion place the lines of a neighboring source at different positions along the trace of the target, which removes much of the confusion (Figure~\ref{fig:cross_config}). Second, they may cover the gap between the two detectors, so that a line falling in the gap of one grating can still be detected in the other. Third, combining an F170LP with an F290LP configuration widens the wavelength coverage, and therefore the redshift range over which the strong lines are accessible.

The confident redshifts are almost all correct in every configuration (Figures~\ref{fig:accuracy}a and \ref{fig:accuracy}b), with only up to two confident but wrong redshifts in each. Among all identified sources (Figures~\ref{fig:accuracy}c and \ref{fig:accuracy}d), the accuracy is 94--97\% for a single configuration and 96--98\% for combinations.

\section{Discussion}
\label{sec:discussion}

\subsection{Failure Cases}
\label{sec:discussion_failures}

Of the $\sim$1800 sources with redshift identifications (Section~\ref{sec:performance}), 47 are incorrect. Only one of them slips into the confident sample, so the other 46 can be flagged by the $P(z)$ criterion alone. Among the 47, only 23 are errors of association, caused by neighboring-line contamination specific to dense-shutter spectroscopy (none confident); 18 associate the lines with the right source but identify them wrongly; and 6 come from limitations of the present algorithm. We discuss each group below.

The 23 errors caused by neighboring lines all occur in sources with a single detectable line (11 cases) or none (12 cases). In 19 cases the contaminant is a line of another source, and in 4 cases an injected false positive. The cross-grating check should have caught them, but 9 contaminants fall at wavelengths that no other grating covers, 11 have SNR $< 7$, below which we do not apply the cross-grating penalty (Section~\ref{sec:algo_priors}), and 3 come from sources one shutter away along the dispersion direction at a similar $y$, which passes the wide wavelength window of the check. A stricter window or visual inspection would remove many of these cases. They amount to 1.3\% of the sample, and none of them is confident.

The 18 interpretation errors correctly associate the lines with the source but misidentify them, and thus give a wrong redshift. Seven, including the confident one discussed in Section~\ref{sec:perf_all}, are lone, faint [\ion{S}{2}], [\ion{N}{2}], [\ion{S}{3}], or \ion{He}{1} lines read as H$\alpha$ or \oiii. Five are single-line sources whose true redshift is an outlier of the photometric redshift, so that the prior favors the wrong identification, such as \oiii\ mistaken for H$\alpha$. The remaining four are two-line degeneracies, [\ion{S}{3}]\,+\,Pa$\beta$ against \oiii\,+\,[\ion{S}{2}], and H$\alpha$\,+\,[\ion{S}{2}]$\lambda6733$ against [\ion{N}{2}]$\lambda6550$\,+\,[\ion{S}{2}]$\lambda6718$, whose wavelength ratios agree to 0.05\% and 0.01\%. We did not apply the photometric prior to these two-line sources, but it would break the degeneracy easily.

The remaining 6 errors come from the current implementation. In five sources the true solution is removed by the pruning of low-ranked candidates that we apply for speed, and only a version contaminated by a false-positive line survives, which is then heavily penalized because that line is absent in the other gratings. A less aggressive pruning would recover these sources, since the true solutions score higher once the penalty is applied. In the last source the lines are identified correctly, but the redshift is off by 0.003, because the polynomial approximation of the spectral trace is less accurate near the edge of the detector.

In summary, the errors introduced by dense-shutter spectroscopy itself are only 1.3\% of the sample, and nearly all of them occur where the cross-grating check happens not to apply. They are of low confidence and easy to exclude, so they affect the sample very little.

\subsection{Limitations of the Current Analysis}
\label{sec:discussion_limitations}

We discuss here the limitations of our current simulation and analysis. First, we do not simulate the signal on the NIRSpec detectors directly. Instead, we simulate the positions and intensities of the emission lines and treat them as a line catalog obtained after the detection. This ignores the spatially varying sky background and the contamination from the continua of the sources. Both are minor in practice: in the dense-shutter observations of \citet{DEugenio2026DarkHorse}, with $\sim$2000 open shutters, the enhanced sky background degrades the line-flux sensitivity by only 30\%, and the source continua stay well below the detector noise. Their exposures are $\sim$10 hr per grating, much deeper than the 2 hr we assume, so the background contributes even less here. We also include only the spatial broadening of the lines, not the broadening by velocity dispersion. Galaxies at $z>3$ typically have velocity dispersions of $\sim$70\kms\ \citep{Turner2017KDS, deGraaff2024msafit}, comparable to the resolution element of the high-resolution gratings and well below that of the medium-resolution ones, so this broadening would only mildly lower the line SNR. Relatedly, we add the same noise to every line, although an extended source spreads over more pixels and is thus measured at a lower SNR. We expect its impact to be mild because extended galaxies are usually luminous and their lines stay well above the detection limit.

Second, the center of the line emission may be offset from the center of the continuum probed by NIRCam. We place the source at its effective center within the shutter and compare the $y$ position of each line with the trace predicted from that center, adopting a positional uncertainty that scales with the SNR and a floor of one fifth of the source in-shutter width (Section~\ref{sec:sim_msa_to_detector}). This holds for most  sources, but an outflow can displace the line emission further and disturb the association. We have not tested these cases, but sources with outflows are usually bright and show several high-ionization lines, so their redshifts should still be determined accurately.

Finally, our Bayesian framework does not yet use all the available information. The first unused quantity is the flux of the galaxy, as a brighter source is expected to have brighter lines. The second is the ratios among the detected lines: we use line ratios only to penalize lines that should have been detected but were not, and not solutions with exotic ratios. The third is the width of a line in the cross-dispersion direction, which usually matches the width of its source within the shutter. All three are commonly used in the visual inspection of spectroscopic redshifts. These expectations, however, do not always hold, and they fail most often for the most interesting sources. A bright source with faint lines may be quenching \citep[e.g.,][]{Strait2023, Looser2024}, while a faint source with bright lines may be an extremely young, metal-poor system \citep[e.g.,][]{Vanzella2023metal}. Exotic line ratios may be nitrogen-rich galaxies \citep[e.g.,][]{Bunker2023GNZ11}, shock-excited gas \citep[e.g.,][]{DEugenio2025shocks}, or high-density systems \citep[e.g.,][]{Trussler2026MNRAS}. Line emission more extended than the continuum may  trace mergers, outflows, or spatially separated quenched and star-forming regions \citep[e.g.,][]{DEugenio2026Sapphires, Zhu2026outflows}. We therefore do not include this information, to keep the association robust for such rare objects.
 
\subsection{Rotation versus Changing the Grating}
\label{sec:discussion_rotation}

Rotating the telescope is an alternative way to move the lines of neighboring sources, and even a small rotation is effective. Sources sharing an MSA row are separated by a median of $9''$, so a rotation of $1^\circ$ displaces them by $0\farcs16$, or 1.6 pixels, in the cross-dispersion direction (right column of Figure~\ref{fig:cross_config}). This is well above the uncertainty of a line centroid, and a contaminating line therefore moves measurably off the trace of the target.

However, rotation changes the target list. In the simulation of Section~\ref{sec:simulation}, rotations of $0.1^\circ$--$1^\circ$ move 46\%--50\% of the sources out of the open shutter areas, almost independently of the angle. Two position angles therefore observe different samples, although together they are more complete, because some sources initially behind the shutter bars can be observed. Changing only the gratings instead keeps every source in the same shutter, so that all of them are observed in every configuration. Rotation also carries a  larger overhead, requiring a new visit with its own slew and target acquisition, while the gratings can be changed within one visit.

\section{Advantages of Dense-Shutter Spectroscopy}
\label{sec:advantages}

Dense-shutter spectroscopy combines the advantages of slitless spectroscopy and conventional NIRSpec MSA observations. Conventionally designed MSA observations are limited by the overlap of spectra on the detector, and therefore target only a small number of sources in the field. NIRCam WFSS disperses every source in the field, but the full sky background falls on the detectors, so its spectra are much shallower. Dense-shutter spectroscopy takes the multiplexing of WFSS while keeping the depth of NIRSpec, in which the shutters block most of the sky background.

Dense-shutter spectroscopy substantially improves the efficiency of surveying redshifts. NIRSpec MSA observations with gratings usually target $\lesssim$200 sources per pointing \citep[e.g.,][]{Bunker2024JADES, CurtisLake2026}. NIRCam WFSS, limited in depth and in wavelength coverage, yields redshifts for only $\sim$50 sources per pointing in a 2 hr exposure \citep[e.g.,][]{Oesch2023, Meyer2024, CoveloPaz2025}: its narrow filter bandwidth covers either H$\alpha$ or the \oiiidoublet\ region of a given source, but rarely both, so most sources are detected in a single line. Dense-shutter spectroscopy instead observes $\sim$2000 sources, secures redshifts for $\sim$1500 of them, and places flux limits on the emission lines of the remaining $\sim$500 (Section~\ref{sec:performance}). Each grating alone produces more than 1000 detected lines (Table~\ref{tab:line_density}), at least twenty times as many as a WFSS pointing. Even counting the 8 hr needed for the four configurations, the yield per hour is still significantly higher, and the data come at more than one spectral resolution. We stress that although all four gratings give only 5\% more sources with a detectable line than the G235M+G395M combination, they yield 7\% more identifications and, crucially for completeness, 16\% more confident redshifts (Table~\ref{tab:performance}).

Dense-shutter spectroscopy  enables large spectroscopic surveys that were previously inaccessible. By efficiently obtaining spectroscopic redshifts for large samples of galaxies, it can deliver samples that are complete to a line-flux limit without preselection of spectroscopic targets. Such samples are particularly valuable for measurements of galaxy clustering, the environmental dependence of galaxy evolution, emission-line luminosity functions, and the evolution of the star-forming galaxy population \citep[e.g.,][]{Kashino2023EIGER, Sun2023ApJSlitless, Shuntov2025}. Their large size also enables statistical studies of chemical enrichment \citep[e.g.,][]{Nakajima2023metallicity, Curti2024metallicity} and increases the likelihood of identifying rare systems, such as extremely metal-poor galaxies \citep[e.g.,][]{Trussler2026}, or galaxies exhibiting Population III signatures \citep[e.g.,][]{Bromm2011ARAA, Vanzella2023metal}. Compared with WFSS, the greater depth and broader wavelength coverage provide access to the faint diagnostic lines needed to constrain metallicity, electron density, ionization state, and the hardness of the ionizing radiation field \citep[e.g.,][]{Arellano2022chemical, Curti2023metallicity, Laseter2024metal}.

Dense-shutter spectroscopy also samples galaxies densely on the sky. Conventional MSA programs rarely target close companions whose spectra would overlap on the detector, causing close pairs and compact groups to be underrepresented in existing spectroscopic samples \citep[e.g.,][]{DEugenio2025DR3}. By obtaining redshifts for essentially all companions, dense-shutter spectroscopy enables measurements of halo dynamical masses \citep{Wu2026}. It can also measure close-pair fractions and merger rates \citep[e.g.,][]{Puskas2025, Puskas2025a}, probe small-scale galaxy clustering and overdensities \citep[e.g.,][]{Harikane2016, Dalmasso2024,Helton2024ApJ,Wu2026overdensity}, and characterize the impact of environment on galaxy evolution \citep[e.g.,][]{Li2025metal}.

\section{Summary}
\label{sec:summary}

We have demonstrated the feasibility of source--line association in NIRSpec dense-shutter spectroscopy at maximum multiplexing, using JADES-based simulations of $\sim$2000 sources per pointing. We use wavelength and cross-dispersion offsets of emission lines, and the change of their relative positions between gratings, to distinguish the lines of different sources. Our Bayesian framework combines these geometric signatures with penalties for physical inconsistency to determine source associations and redshifts jointly.

Combining four gratings, we obtain redshift solutions for $\sim$1500 sources per pointing with 99.9\% accuracy, and for a total of $\sim$1800 sources with 98\% accuracy. Errors caused by contaminating lines account for only 1.3\% of the sample, and none of them is classified as confident. Spectral overlap therefore does not compromise the reliability of the  sample.

These results demonstrate the potential of dense-shutter spectroscopy that combines high multiplexing with the sensitivity and broad wavelength coverage of NIRSpec. Dense-shutter spectroscopy can substantially increase survey efficiency while reducing target-selection biases imposed by spectral overlap. This strategy opens a new avenue for large spectroscopic surveys, enabling large-scale studies of galaxy clustering, chemical enrichment, environmental effects, and halo dynamical masses.

\begin{acknowledgements}

This work is based on observations made with the NASA/ESA/CSA James Webb Space Telescope. The data were obtained from the Mikulski Archive for Space Telescopes at the Space Telescope Science Institute, which is operated by the Association of Universities for Research in Astronomy, Inc., under NASA contract NAS 5-03127 for JWST.

Z.W. and D.J.E. are supported by the Simons Foundation Investigator program. D.J.E., B.D.J., B.E.R., Z.J., and C.N.A.W. acknowledge support from the NIRCam Science Team contract to the University of Arizona, NAS5-02105. D.J.E. is also supported by NASA through a grant from the Space Telescope Science Institute, which is operated by the Association of Universities for Research in Astronomy, Inc., under NASA contract NAS5-03127. B.E.R. also acknowledges support from JWST Program 3215. F.D'E. acknowledges support by the Science and Technology Facilities Council (STFC), by the ERC through Advanced Grant 695671 ``QUENCH'', and by the UKRI Frontier Research grant RISEandFALL. A.J.B. acknowledges funding from the ``FirstGalaxies'' Advanced Grant from the European Research Council (ERC) under the European Union's Horizon 2020 research and innovation program (Grant agreement No. 789056). P.R. acknowledges support from the University of Texas at Austin Cosmic Frontier Center. S.T. acknowledges support by the Royal Society Research Grant G125142. J.W. gratefully acknowledges support from the Cosmic Dawn Center through the DAWN Fellowship; the Cosmic Dawn Center (DAWN) is funded by the Danish National Research Foundation under grant No.~140.

We acknowledge the use of Claude Code (Anthropic) and Codex (OpenAI) for assistance in developing the analysis code and for suggesting improvements to the clarity and flow of the text. The authors have reviewed all content and take full responsibility for the scientific results and conclusions.

\end{acknowledgements}

\appendix
\twocolumngrid
\restartappendixnumbering

\section{Line Displacement between Gratings}
\label{app:relative_line_positions}

\begin{figure}
	\centering
	\includegraphics[width=0.8\linewidth]{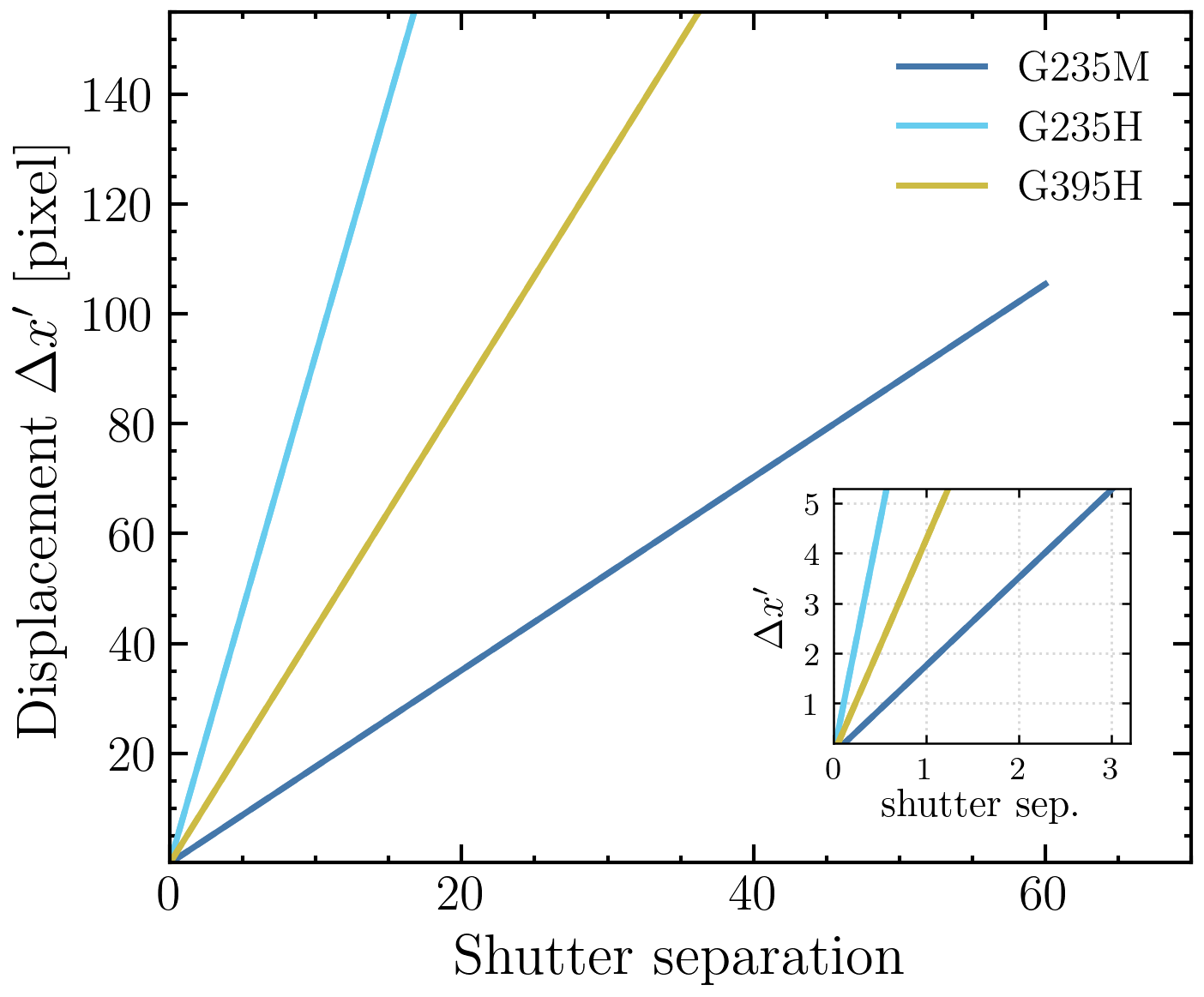}
	\caption{Changes in the relative positions of emission lines between gratings. We choose two lines from different sources in the same MSA row that fall at the same detector $x$ in G395M. The curves show their displacement $\Delta x$ in G235M, G235H, and G395H as a function of the shutter separation of the two sources, in units of the shutter pitch, computed with the full NIRSpec trace model of Section~\ref{sec:sim_msa_to_detector}. The inset zooms in on separations of a few shutters.
\label{fig:displacement}}
\end{figure}

Here we consider the problem: when a line of source $s_2$ is misidentified as the line of wavelength $\lambda_1$ of source $s_1$, how far does it lie from the position expected for that line once the resolving power changes from $R$ to $R'$?

Equivalently, consider two lines of observed wavelengths $\lambda_1$ and $\lambda_2$, belonging to $s_1$ and $s_2$ respectively, that coincide on the detector when observed with the grating of resolving power $R$. We derive their offset when the same two lines are observed with $R'$.

Let $x_{{\rm s},1}$ and $x_{{\rm s},2}$ denote the positions of the two sources along the dispersion direction, projected into detector-pixel units. The MSA shutter width (center to center) is 0.27$''$, while the NIRSpec detector pixel size is 0.1$''$, so $x_{{\rm s},1}-x_{{\rm s},2}=2.7n$ pixels when the two sources are $n$ shutters away.

 We approximate the dispersion as constant over the wavelength interval of interest for simplicity. The detector positions of two emission lines at observed wavelengths $\lambda_1$ and $\lambda_2$ are
 
\begin{align}
    x_i &= x_{{\rm s},i} + D(\lambda_i-\lambda_0) + c_0, \\
    x_i' &= x_{{\rm s},i} + D'(\lambda_i-\lambda_0) + c_0',
\end{align}

\noindent
where $i=1,2$, $D \equiv dx/d\lambda$ and $D'$ are the dispersions of the two gratings, $\lambda_0$ is a reference wavelength, and $c_0$ and $c_0'$ are grating-dependent offsets.

Defining

\begin{equation}
    \Delta x_{\rm s} \equiv x_{{\rm s},2}-x_{{\rm s},1},
    \qquad
    \Delta\lambda \equiv \lambda_2-\lambda_1,
\end{equation}

\noindent
the line separations in the two gratings are

\begin{align}
\label{eq:dx}
    \Delta x &\equiv x_2-x_1
        = \Delta x_{\rm s} + D\Delta\lambda, \\
    \Delta x' &\equiv x_2'-x_1'
        = \Delta x_{\rm s} + D'\Delta\lambda.
\end{align}

\noindent
The grating-dependent offsets cancel, giving

\begin{equation}
    \Delta x'-\Delta x = (D'-D)\Delta\lambda.
\label{eq:dx_diff}
\end{equation}

When the two lines overlap in the first grating, we have $\Delta x=0$. Combining Equations~\ref{eq:dx} and \ref{eq:dx_diff}, the separation in the second grating is

\begin{equation}
    \Delta x' \simeq -\Delta x_{\rm s}\left({R'}/{R}-1\right),
    \label{eq:overlapping_line_separation}
\end{equation}

\noindent
where we have used the resolving power
$R\propto D$. Thus, lines from different sources that coincide in one grating generally separate in another. The induced separation increases with both the projected source separation and the fractional change in dispersion.

For illustration, with a projected source separation of $|\Delta x_{\rm s}|\simeq 2.7$~pixels for adjacent shutters, a dispersion ratio of $D'/D\simeq 3$ between high- and medium-resolution gratings produces $|\Delta x'|\simeq 5.4$~pixels for lines that initially overlap. Even two medium-resolution gratings provide useful leverage: taking G395M as the first grating and G235M as the second, $D'/D\simeq 1.6$ gives $|\Delta x'|\simeq 1.6$~pixels. Such a separation can constrain source associations.

The effect is substantially larger for more widely separated sources. For example, sources separated by 100 shutter pitches have $|\Delta x_{\rm s}|\simeq 270$~pixels, yielding $|\Delta x'|\simeq 160$~pixels when switching from G395M to G235M for lines that overlap in G395M.

Figure~\ref{fig:displacement} shows this effect computed with the full trace model of Section~\ref{sec:sim_msa_to_detector}. Source~1 sits in a shutter of MSA quadrant~3 with a line at $\lambda_1=3.15\,\mu$m, and source~2 lies in the same shutter row with its line wavelength $\lambda_2$ chosen so that the two lines land on the same detector column in G395M. The displacement grows almost linearly with the shutter separation, at 1.8, 4.3, and 9.2~pixels per shutter in G235M, G395H, and G235H, respectively, in agreement with Equation~\ref{eq:overlapping_line_separation} evaluated with the local dispersion ratios of these gratings. Even between the two medium-resolution gratings, G395M and G235M, a separation of a single shutter already displaces the two lines by about 2~pixels, larger than the typical uncertainty of a line centroid.

\section{Empirical Line-Ratio Distributions}
\label{app:line_ratios}

\begin{figure*}
\centering
\includegraphics[width=0.95\textwidth]{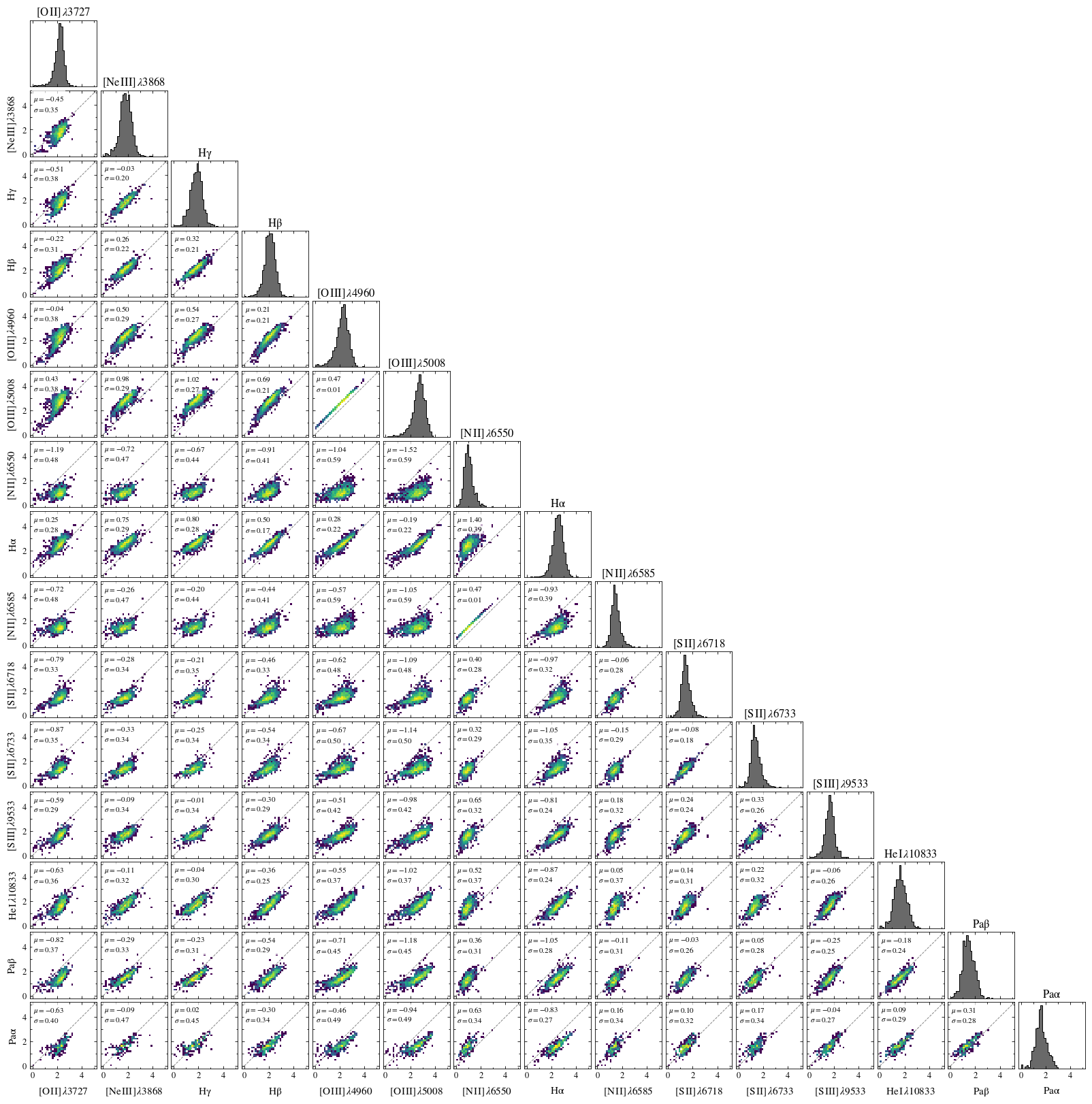}
\caption{Pairwise flux distributions of the 15 reference emission lines measured from the JADES DR4 NIRSpec spectra.
Diagonal panels show the distribution of the logarithmic flux of each line; off-diagonal panels show the joint logarithmic flux distributions of each line pair, with the dashed line marking equal fluxes.
The annotated $\mu$ and $\sigma$ are the median and standard deviation of the logarithmic flux ratio of the vertical-axis line to the horizontal-axis line.
\label{fig:line_ratios}}
\end{figure*}

Figure~\ref{fig:line_ratios} shows the joint flux distributions of the 15 reference emission lines measured from the JADES DR4 NIRSpec spectra, which we use to simulate missing line fluxes (Section~\ref{sec:sim_catalog}) and as the line-ratio priors of the Bayesian framework (Section~\ref{sec:algorithm}).

\bibliographystyle{aasjournalv7}
\bibliography{reference}

\end{document}